\documentclass[12pt]{article}
\usepackage{amsmath}
\usepackage{amssymb}
\usepackage{epsf}
\DeclareSymbolFont{boldletters}{OML}{cmm} {b}{it}
\DeclareSymbolFontAlphabet{\mathbit}{boldletters}
\DeclareMathSymbol{\alpha}{\mathalpha}{letters}{"0B}
\DeclareMathSymbol{\beta}{\mathalpha}{letters}{"0C}
\DeclareMathSymbol{\gamma}{\mathalpha}{letters}{"0D}
\DeclareMathSymbol{\delta}{\mathalpha}{letters}{"0E}
\DeclareMathSymbol{\epsilon}{\mathalpha}{letters}{"0F}
\DeclareMathSymbol{\zeta}{\mathalpha}{letters}{"10}
\DeclareMathSymbol{\eta}{\mathalpha}{letters}{"11}
\DeclareMathSymbol{\theta}{\mathalpha}{letters}{"12}
\DeclareMathSymbol{\iota}{\mathalpha}{letters}{"13}
\DeclareMathSymbol{\kappa}{\mathalpha}{letters}{"14}
\DeclareMathSymbol{\lambda}{\mathalpha}{letters}{"15}
\DeclareMathSymbol{\mu}{\mathalpha}{letters}{"16}
\DeclareMathSymbol{\nu}{\mathalpha}{letters}{"17}
\DeclareMathSymbol{\xi}{\mathalpha}{letters}{"18}
\DeclareMathSymbol{\pi}{\mathalpha}{letters}{"19}
\DeclareMathSymbol{\rho}{\mathalpha}{letters}{"1A}
\DeclareMathSymbol{\sigma}{\mathalpha}{letters}{"1B}
\DeclareMathSymbol{\tau}{\mathalpha}{letters}{"1C}
\DeclareMathSymbol{\upsilon}{\mathalpha}{letters}{"1D}
\DeclareMathSymbol{\phi}{\mathalpha}{letters}{"1E}
\DeclareMathSymbol{\chi}{\mathalpha}{letters}{"1F}
\DeclareMathSymbol{\psi}{\mathalpha}{letters}{"20}
\DeclareMathSymbol{\omega}{\mathalpha}{letters}{"21}
\DeclareMathSymbol{\varepsilon}{\mathalpha}{letters}{"22}
\DeclareMathSymbol{\vartheta}{\mathalpha}{letters}{"23}
\DeclareMathSymbol{\varpi}{\mathalpha}{letters}{"24}
\DeclareMathSymbol{\varrho}{\mathalpha}{letters}{"25}
\DeclareMathSymbol{\varsigma}{\mathalpha}{letters}{"26}
\DeclareMathSymbol{\varphi}{\mathalpha}{letters}{"27}
\DeclareMathSymbol{\Gamma}{\mathalpha}{letters}{"00}
\DeclareMathSymbol{\Delta}{\mathalpha}{letters}{"01}
\DeclareMathSymbol{\Theta}{\mathalpha}{letters}{"02}
\DeclareMathSymbol{\Lambda}{\mathalpha}{letters}{"03}
\DeclareMathSymbol{\Xi}{\mathalpha}{letters}{"04}
\DeclareMathSymbol{\Pi}{\mathalpha}{letters}{"05}
\DeclareMathSymbol{\Sigma}{\mathalpha}{letters}{"06}
\DeclareMathSymbol{\Upsilon}{\mathalpha}{letters}{"07}
\DeclareMathSymbol{\Phi}{\mathalpha}{letters}{"08}
\DeclareMathSymbol{\Psi}{\mathalpha}{letters}{"09}
\DeclareMathSymbol{\Omega}{\mathalpha}{letters}{"0A}
\newcommand{\mbit}[1]{{\mathbit#1}}

\begin{document}

\title{
\begin{flushright}
\small{
KYUSHU-HET-76 }
\end{flushright}
\vspace{2cm}
An asymptotic formula for models with caustics}%title here

\author{
Tomohiko~Sakaguchi\thanks{tomohiko@higgs.phys.kyushu-u.ac.jp}
\\
Department of Physics, Kyushu University, Fukuoka 812-8581, Japan\\\\}

\date{\today}

\maketitle

\abstract{We introduce an asymptotic formula for calculating 
quantum mechanical and
quantum theoretical models with caustics, 
like the Nambu--Jona-Lasinio(NJL) model.
This asymptotic formula is given by the form attached the extra term, which
suppresses the divergence induced because of caustics, 
to the leading term of the WKB approximation.
This formula guarantees validity of the mean field
 approximation of models with caustics.}

\maketitle\thispagestyle{empty}
\newpage
%%%%%%%%%%%%%%%%%%%%%%%%%%%%%%%%%%%%%%%%%%%
\section{Introduction}
%%%%%%%%%%%%%%%%%%%%%%%%%%%%%%%%%%%%%%%%%%%

There is the WKB approximation method as one of 
the nonperturbative approximation methods:
the action is expanded around the classical solution, and terms 
with derivatives of greater than third order are treated as
perturbations, to obtain a series expansion expressed 
in terms of a loop expansion parameter, such as 
$\hbar$ or $1/N$ \cite{rf:KOS}.
Then it is very important that 
positivity for all eigenvalues of a second derivative of the 
action at the classical solution, 
that is, the convexity condition of the 
action must be satisfied.
When this approximation method is utilized to 
the Nambu--Jona-Lasinio(NJL) model,
which is 
one of the most famous field theoretical models exhibiting 
dynamical chiral symmetry breaking phenomena, 
many interesting results are made in a level of the mean field
approximation.
For example, T.~Inagaki $\textit{et al.}$~\cite{rf:IKM} 
have investigated the phase structure of the NJL model 
at finite temperature and chemical potential 
in an external magnetic field and gained the interesting results such as 
the magnetic catalysis.

Nevertheless the mean field approximation in the NJL model is open to
question:
for example,
in reference~\cite{rf:KS}, we had calculated the contribution of 
one loop diagrams of the auxiliary fields for 
the the effective potential 
and Gap equation of the NJL model. 
Then the infrared divergence occurs from the loop diagram of the
massless auxiliary field.
This fact implies that 
we always have to introduce a current mass 
in order to suppress this infrared divergence, and either do the mean
field approximation.
Otherwise the WKB approximation fails because the convexity condition 
is not satisfied.
This discourages the interesting results obtained in the chiral limit.
Therefore it is important 
to investigate whether these mean field approximations
is justified in terms of the asymptotic expansion.

Generally when one and more eigenvalues
of the second derivative of action vanish at the classical solution,
higher order terms of the WKB approximation diverge. 
This situation is often called `caustics'. 
L.~S.~Shulman~\cite{rf:S} says that this divergence occurs approximately 
because the third derivative of action does not vanish 
at the classical solution.
In fact, he has given the $\textit{ad hoc}$ treatment of caustics 
for a Green function in quantum mechanics.
Unfortunately, This method cannot apply to calculate the effective
potential for the NJL model in the chiral limit.
We need to consider other method for the treatment of caustics.

It is well-known that similar circumstance like caustics occurs 
when we calculate an asymptotic expansion
of an integral using the saddle point method~\cite{rf:C}~\cite{rf:O}.
For example, let us consider the Bessel function
\begin{eqnarray}
 J_N (N \alpha) = \frac{1}{2 \pi i} \int_{\infty - \pi i}^{\infty + \pi i}
e^{N(\alpha \sinh z -  z)} dz \,\,\, (\mbox{Re} \, \alpha > 0) \ , 
\label{Bessel}  
\end{eqnarray} 
where Re $\alpha$ stands for the real part of $\alpha$.
If $\alpha = 1$, the second derivative of the exponent 
becomes zero at the saddle point $z = 0$. 
In such a case, there are two $\alpha$-dependent 
and nearly coincident saddle points
in the neighborhood of $\alpha=1$. 
Utilizing the two saddle points and 
replacing the exponent $\alpha \sinh z -  z$ 
with a cubic function with respect to a new variable,
C.~Chester~$\textit{et al.}$  
have presented the method giving a uniform asymptotic expansion 
as $\alpha \to 1$ and $N \to \infty$
\cite{rf:CFU} \cite{rf:F} \cite{rf:U}.
The application for an integral with multi-variables had been presented 
by N.~Bleistein~\cite{rf:B}.

We would like to apply their method to the NJL model.
However this method cannot use directly to the NJL model, 
because the classical equation of the NJL model 
has only one classical solution
or at least we do not know whether the another solution,
which coincides with the classical solution in the chiral limit, exists.  
We need to make the method that we can use for the model having
only one classical solution so that we can obtain the same property 
as the asymptotic expansion presented by C.~Chester $\textit{et al.}$.

In reference~\cite{rf:KS2}, an other method obtaining 
an asymptotic expansion of an integral 
which the second derivative of the exponent of the integrand 
becomes zero at the saddle point is introduced.
Although this method does not give a uniform asymptotic expansion,
it needs only one expansion point and 
the result coincides with the one based on the method
presented by C.~Chester $\textit{et al.}$ in the neighborhood of
caustics. 
It is the most remarkable thing in this method that 
this expansion point always exists 
even if we do not know whether the another saddle point exists.
Utilizing this fact,
we introduce other approach for obtaining the asymptotic formula 
of the models with caustics.

This paper is organized as follows: 
in section 2, 
we mention the method obtaining  
the asymptotic formula of an integral with one variable,
which had been utilized in the paper~\cite{rf:KS2}. 
This method utilizes a point at which the second derivative of the exponent 
becomes zero as an expansion point of the exponent, 
rather than the saddle point. 
Then we rewrite the obtained formula with respect to the saddle point
and show the formula is the form attached the extra term which
suppresses the divergence to the result of the saddle point method.
In the end of section 2, we show that the formula coincides with the
leading order of the asymptotic expansion given by  
C.~Chester $\textit{et al.}$ in the neighborhood of caustics.  
In section 3, we apply this method to obtain  
an asymptotic formula of an integral with multi-variables.
In the same way as the result of section 2,
We show that the formula is the form attached the extra term which
suppresses the divergence to the result of the WKB approximation.
This implies that the mean field approximation of this formula
coincides with the one of the WKB approximation
and gives the important result that  
the mean field approximation of the model with caustics, 
like the NJL model, 
is justified in the sense of the formula whose result converges. 
The final section is devoted to conclusions and discussions.

\section{An asymptotic formula of an integral with one variable}

In this section, we establish 
an asymptotic formula as $N \to \infty$ of an integral
\begin{eqnarray} 
 I(\alpha,N) = \int_{c} g(z) e^{N f(z,\alpha)} d z, \label{integral}
\end{eqnarray}
where the real functions $g(z), f(z,\alpha)$ are analytic functions 
of their arguments, $N$ is a large positive parameter and $c$ is some
contour in the complex $z$ plane.
The function $f(z,\alpha)$ has the saddle point $z_0$ which 
is dependent on $\alpha$. 
The second derivative on the steepest descent,
$\partial^2 f(z,\alpha)/ \partial z^2 |_{z=z_0(\alpha)}$, becomes zero 
at some value of $\alpha$, say $\hat{\alpha}$, and takes negative value
for $\alpha > \hat{\alpha}$.\footnote{In the case of the integral with
one variable, it is expected that 
the another saddle point, say $z_0^*$, 
which coincides with the saddle point $z_0$ at
$\alpha = \hat{\alpha}$ always exists. 
In this paper, as we would like to derive the method utilized 
only one saddle point,
we assume that we cannot deform the integration path 
so that it traces the steepest descent of the point $z_0^*$ 
when $\alpha > \hat{\alpha}$ and we do not utilize the point $z_0^*$. 
Also, we do not consider about 
$\alpha < \hat{\alpha}$ because the saddle points $z_0$ and $z_0^*$ 
become the complex conjugate each other and it is out of our aim.}

We expand $f(z,\alpha)$, $g(z)$ around $\tilde{z}$ which satisfy
with $f^{(2)}(\tilde{z},\alpha) = 0$ 
(and  
$ f^{(1)}(\tilde{z},\hat{\alpha}) = 0$):
\begin{eqnarray}
 I(\alpha,N) &\sim& \int_c \left( g(\tilde{z}) + O(z-\tilde{z}) \right) 
 \nonumber \\
 &\times& \exp \Bigg[ N \Bigg\{f(\tilde{z},\alpha) + 
  f^{(1)}(\tilde{z},\alpha)(z-\tilde{z}) 
  + \frac{1}{3!} f^{(3)}(\tilde{z},\alpha)(z-\tilde{z})^3 
  + O((z-\tilde{z})^4)
   \Bigg\} \Bigg] d z \ ,
  \nonumber \\
\label{expand} 
\end{eqnarray}
where we utilized the abbreviations 
$f^{(n)}(z,\alpha) = \partial^n f(z,\alpha)/\partial z^n$.
We assume that $f^{(3)}(\tilde{z},\hat{\alpha}) \neq 0$. 
Considering only in the neighborhood of $\hat{\alpha}$,
we can regard $f^{(1)}(\tilde{z},\alpha)$ as very small quantity,
especially, 
we set that $|f^{(1)}(\tilde{z},\alpha)| \sim O (1/N^{\frac{2}{3}})$ 
for sufficiently large $N$. 
So we can consider both terms $f^{(1)}(\tilde{z},\alpha)(z-\tilde{z})$ and 
$f^{(3)}(\tilde{z},\alpha)(z-\tilde{z})^3$ as same order terms,
in the neighborhood of $\tilde{z}$ on the $z$ space.
We retain terms, $g(\tilde{z})$, $f(\tilde{z},\alpha)$, 
$f^{(1)}(\tilde{z},\alpha)(z-\tilde{z})$ and 
$f^{(3)}(\tilde{z},\alpha)(z-\tilde{z})^3$ as the principle part,
and remaining terms with higher derivatives are treated as
negligible quantities, to obtain 
\begin{eqnarray}
I(\alpha,N) &\sim&
 g(\tilde{z}) \left( \frac{2}{f^{(3)}(\tilde{z},\alpha)} 
		   \right)^{\frac{1}{3}} 
 \frac{e^{N f(\tilde{z},\alpha)}}{N^{\frac{1}{3}}}
 \int_{c_i} d t \,\, e^{\frac{1}{3}t^3 - \zeta' t} \ ,
\label{expand3}
\end{eqnarray}
where we have set
\begin{eqnarray}
 \left(\frac{f^{(3)}(\tilde{z},\alpha)}{2} \right)^{\frac{1}{3}}(z-\tilde{z})
  \equiv \frac{t}{N^{\frac{1}{3}}}, \,\,\, 
  - \left(\frac{2}{f^{(3)}(\tilde{z},\alpha)} \right)^{\frac{1}{3}} 
  f^{(1)}(\tilde{z},\alpha) \equiv \zeta (\alpha) \  \label{tdef}
\end{eqnarray}
and $\zeta' = N^{\frac{2}{3}} \zeta$. 
The integral part of Eq.~(\ref{expand3}) just corresponds to
the Airy integral 
\begin{eqnarray}
 \mbox{Ai}(\zeta') = \frac{1}{2 \pi i} \int_{c_A} dt 
  \,\, e^{\frac{1}{3}t^3-\zeta' t } \ .  
  \label{Air} 
\end{eqnarray}
Therefore we can obtain the asymptotic formula
\begin{eqnarray}
 I(\alpha,N) &\sim& 
(2 \pi i) g(\tilde{z})
\left( \frac{2}{f^{(3)}(\tilde{z},\alpha)} 
		   \right)^{\frac{1}{3}} 
 \frac{e^{N f(\tilde{z},\alpha)}}{N^{\frac{1}{3}}} \mbox{Ai}(\zeta') \ .
\label{expand2}
\end{eqnarray}

We can rewrite the asymptotic formula~(\ref{expand2}) 
with the saddle point $z_0$, 
not the point $\tilde{z}$.
When $\alpha$ is in a neighborhood of $\hat{\alpha}$, 
a distance between $\tilde{z}$ and $z_0$, $|\tilde{z}-z_0|$, 
becomes very small.
So we can expand $g(\tilde{z})$, $f(\tilde{z},\alpha)$ and its derivatives 
at the point $\tilde{z}$
around the saddle point $z_0$:
\begin{eqnarray}
g(\tilde{z}) &=& g(z_0) + \cdots , \label{g0} \\
 f(\tilde{z},\alpha) &=& f(z_0,\alpha) + \frac{1}{2} f^{(2)} (z_0,\alpha) 
(\tilde{z} - z_0)^2 + \frac{1}{3!} f^{(3)} (z_0,\alpha) 
(\tilde{z} - z_0)^3 + \cdots \ , \label{f0} \\ 
 f^{(1)}(\tilde{z},\alpha) &=& f^{(2)}(z_0,\alpha) (\tilde{z}-z_0) 
+ \frac{1}{2} f^{(3)} (z_0,\alpha) (\tilde{z} - z_0)^2 + \cdots \ , 
\label{f1} \\  
0 = f^{(2)}(\tilde{z},\alpha) &=& f^{(2)}(z_0,\alpha) 
+ f^{(3)} (z_0,\alpha) (\tilde{z} - z_0) + \cdots \ ,  \label{f2}\\  
f^{(3)} (\tilde{z},\alpha) &=& 
f^{(3)} (z_0,\alpha) + \cdots \ . \label{f3}
\end{eqnarray}   
If we assume that $f^{(2)}(z_0,\alpha) \sim O( 1/N^{\frac{1}{3}} )$
and that $N$ is sufficiently large, 
Eq.~(\ref{f2}) gives
\begin{eqnarray}
 \tilde{z}-z_0 \approx \frac{-f^{(2)}(z_0,\alpha)}{f^{(3)}(z_0,\alpha)} 
\sim O \left( \frac{1}{N^{\frac{1}{3}}} \right) \ ,
\end{eqnarray}
and Eqs.~(\ref{g0}), (\ref{f0}), (\ref{f1}) and (\ref{f3}) 
are approximately given by 
\begin{eqnarray}
g(\tilde{z}) &\approx& g(z_0) \ , \label{g0tilde}\\
f(\tilde{z},\alpha) &\approx& f(z_0,\alpha) 
- \frac{1}{3} \frac{(-f^{(2)}(z_0,\alpha))^3}{(f^{(3)}(z_0,\alpha))^2} \ , \\
f^{(1)}(\tilde{z},\alpha) &\approx& - \frac{1}{2} 
\frac{(-f^{(2)}(z_0,\alpha))^2}{f^{(3)}(z_0,\alpha)} \sim 
O \left( \frac{1}{N^{\frac{2}{3}}} \right) \ , \label{f1tilde} \\
f^{(3)}(\tilde{z},\alpha) &\approx& f^{(3)} (z_0,\alpha) \label{f3tilde}\ ,
\end{eqnarray}
respectively. 
Substituting Eqs.~(\ref{g0tilde})-(\ref{f3tilde}) into Eq.~(\ref{expand2}),
We obtain the asymptotic formula of the integral~(\ref{integral}) 
in terms of the saddle point $z_0$:
\begin{eqnarray}
 I(\alpha,N) &\sim& 
(2 \pi i) g(z_0)
\left( \frac{2}{f^{(3)}(z_0, \alpha)} \right)^{\frac{1}{3}} 
 \frac{e^{N f(z_0,\alpha)- \frac{2}{3} \zeta^{' \frac{3}{2}} (\alpha)}}
{N^{\frac{1}{3}}} 
\mbox{Ai}(\zeta') \nonumber \\
&=& g(z_0) e^{N f(z_0,\alpha)} \sqrt{\frac{2 \pi}{-N f^{(2)}(z_0,\alpha)}}
(2 \sqrt{\pi} i) \zeta^{'\frac{1}{4}} e^{-\frac{2}{3} \zeta^{'\frac{3}{2}}}
\mbox{Ai}(\zeta') \ ,
\label{asymptotic}
\end{eqnarray}
and $\zeta(\alpha)$ is deformed as 
\begin{eqnarray} 
 \zeta (\alpha) \equiv  \left[\frac{(-f^{(2)}(z_0,\alpha))^3}
{2 (f^{(3)}(z_0,\alpha))^2} \right]^{\frac{2}{3}}  \ . \label{zetafun}
\end{eqnarray}
The form of the asymptotic formula~(\ref{asymptotic}) is 
the form attached the extra term 
to the leading order of the asymptotic expansion based on the saddle
point method, so that the divergence due to the term
$f^{(2)}(z_0,\alpha)$ at $\alpha=\hat{\alpha}$ is suppressed.  
Expanding the Airy function~(\ref{Air}) by use of the
saddle point method as follows
\begin{eqnarray} 
 \mbox{Ai}(\zeta') \sim \frac{N^{\frac{1}{3}}}{2 \pi i} e^{ \frac{2}{3} 
\zeta^{' \frac{3}{2}}} \sqrt{\frac{\pi}{N \zeta^{\frac{1}{2}}}} 
= \frac{e^{\frac{2}{3} \zeta^{'\frac{3}{2}}}}
{(2 \sqrt{\pi} i) \zeta^{'\frac{1}{4}}} \ , 
\end{eqnarray}
and substituting this expansion in the formula~(\ref{asymptotic}) takes off
the extra term and leaves 
the leading term of the asymptotic expansion of the 
integral~(\ref{integral}) by the saddle point method.
We find that, as in Schulman's textbook~\cite{rf:S},
the term $f^{(3)}(z_0,\alpha)$ 
plays a important role in order to
converge the result of the WKB approximation.

Finally, we show that the asymptotic formula~(\ref{asymptotic})
coincides with the uniform asymptotic expansion of the 
integral~(\ref{integral}) 
based on 
the method presented by C.~Chester~$\textit{et al.}$ 
in the neighborhood of $\hat{\alpha}$.
The leading order of the asymptotic expansion due to their method 
is given as
\begin{eqnarray}
 I(\alpha,N) &\sim& 
(2 \pi i) a_0(\alpha) 
 \frac{e^{N A(\alpha)}}{N^{\frac{1}{3}}} \mbox{Ai}(\zeta') \ , \label{Chester}
\label{expand2}
\end{eqnarray}
where 
\begin{eqnarray}
 A (\alpha) &=& \frac{1}{2}(f(z_0,\alpha)+f(z_0^*,\alpha))  \ , \label{A*} \\
  \zeta (\alpha) &=& \left[\frac{3}{4} (f(z_0,\alpha)-f(z_0^*,\alpha))  
			 \right]^{\frac{2}{3}} \ ,  \label{zeta*}\\
  a_0 (\alpha) &=& 
  \frac{1}{2} \left( 
    g(z_0) \sqrt{ \frac{2 \zeta^{\frac{1}{2}}}{-f^{(2)}(z_0,\alpha)}}
   +g(z_0^*) \sqrt{\frac{2 \zeta^{\frac{1}{2}}}{f^{(2)}(z_0^*,\alpha)}}    
	 \right) \ . \label{a0*}
\end{eqnarray}
A point $z_0^*$ represents the another saddle point 
that coincides with the saddle point $z_0$ 
when $\alpha$ approaches to $\hat{\alpha}$.\footnote{In making 
change of variables, the saddle points $z_0$ and $z_0^*$
are corresponded to $-\zeta^{\frac{1}{2}}$ and $\zeta^{\frac{1}{2}}$, 
respectively.}     
When $\alpha$ is in a neighborhood of $\hat{\alpha}$, 
a distance between $z_0^*$ and $z_0$, $|z_0^*-z_0|$, 
becomes very small.
Therefore we can expand $g(z_0^*)$, $f(z_0^*,\alpha)$ and its derivatives 
at the another saddle point $z_0^*$
around the saddle point $z_0$:
\begin{eqnarray}
g(z_0^*) &=& g(z_0) + \cdots \ , \label{g0*} \\ 
f(z_0^*,\alpha) &=& f(z_0,\alpha) + \frac{1}{2} f^{(2)} (z_0,\alpha) 
(z_0^* - z_0)^2 + \frac{1}{3!} f^{(3)} (z_0,\alpha) 
(z_0^* - z_0)^3 + \cdots \ , \label{f0*} \\ 
 0 = f^{(1)}(z_0^*,\alpha) &=& f^{(2)}(z_0,\alpha) (z_0^*-z_0) 
+ \frac{1}{2} f^{(3)} (z_0,\alpha) (z_0^*- z_0)^2 + \cdots \ , 
\label{f1*} \\  
 f^{(2)}(z_0^*,\alpha) &=& f^{(2)}(z_0,\alpha) 
+ f^{(3)} (z_0,\alpha) (z_0^* - z_0) + \cdots \ ,  \label{f2*}\\  
f^{(3)} (z_0^*,\alpha) &=&  f^{(3)} (z_0,\alpha) + \cdots \ . \label{f3*}
\end{eqnarray}   
If we assume that $f^{(2)}(z_0,\alpha) \sim O( 1/N^{\frac{1}{3}} )$
and that $N$ is sufficiently large, 
Eq.~(\ref{f1*}) gives
\begin{eqnarray}
 z_0^*-z_0 \approx - \frac{2 f^{(2)}(z_0,\alpha)}{f^{(3)}(z_0,\alpha)} 
\sim O \left( \frac{1}{N^{\frac{1}{3}}} \right) \ ,
\end{eqnarray}
and Eqs.~(\ref{g0*}), (\ref{f0*}), (\ref{f2*}) and (\ref{f3*}) 
are approximately given by 
\begin{eqnarray}
 f(z_0^*,\alpha) &\approx& f(z_0,\alpha) 
 - \frac{2}{3} \frac{(-f^{(2)}(z_0,\alpha))^3}{(f^{(3)}(z_0,\alpha))^2} \ ,
 \label{f0tilde*} \\
 f^{(2)}(z_0^*,\alpha) &\approx& - f^{(2)}(z_0,\alpha) \ , \label{f2tilde*} \\
 f^{(3)}(z_0^*,\alpha) &\approx& f^{(3)} (z_0,\alpha) \ , \label{f3tilde*}
\end{eqnarray}
respectively. 
Substituting Eqs.~(\ref{f0tilde*})-(\ref{f3tilde*}) into
Eqs~(\ref{A*}), (\ref{zeta*}) and (\ref{a0*}), 
we obtain 
\begin{eqnarray}
 A(\alpha) &\approx& f(z_0,\alpha) 
- \frac{1}{3}  \frac{(-f^{(2)}(z_0,\alpha))^3}{(f^{(3)}(z_0,\alpha))^2} \ , 
\,\,\,
\zeta (\alpha) \approx \left[ 
\frac{1}{2}  \frac{(-f^{(2)}(z_0,\alpha))^3}{(f^{(3)}(z_0,\alpha))^2}  
\right]^{\frac{2}{3}} \ , \\ 
a_0 (\alpha) &\approx& g(z_0) 
\left( \frac{2 \zeta^{\frac{1}{2}}}{-f^{(2)}(z_0,\alpha)} 
\right)^{\frac{1}{2}} \ ,
\end{eqnarray}
and substitution of these expressions 
in the asymptotic expansion~(\ref{Chester}) 
gives the asymptotic formula~(\ref{asymptotic}). 
This fact shows that, even if we do not know 
whether the another saddle point exists, 
we can derive the asymptotic formula~(\ref{asymptotic}) consistent with 
the result of C.~Chester~$\textit{et al.}$ at least in the neighborhood
of $\hat{\alpha}$.

\section{An asymptotic formula of an integral with multi-variables}

Next we investigate an asymptotic formula of an integral with
$n$ component variables, 
which is formed as 
\begin{eqnarray}
I_n (\alpha, N) = \int_D e^{N F(\mbit{x},\alpha)} d^n x \ , \label{Integral} 
\end{eqnarray}
where $\mbit{x}=(x_1, x_2 , \cdots , x_n) \ ,
 \ d^n x = d x_1 dx_2 \cdots dx_n $ 
and $D$ is a region in $n$ dimensional space.
A saddle point of $F(\mbit{x},\alpha)$ is expressed as  
\begin{eqnarray}  
 \mbit{x}_0(\alpha) = 
  (x_{01}(\alpha) , x_{02}(\alpha) , \cdots , x_{0n}(\alpha)) \ .
\end{eqnarray}
Without loss of generality an expansion of $F(\mbit{x},\alpha)$ 
around the saddle point $\mbit{x}_0$ can be given by 
\begin{eqnarray}
 F(\mbit{x},\alpha) &=& F(\mbit{x}_0,\alpha) +\frac{1}{2} \sum_{i=1}^n 
\lambda_i (\alpha) (x_i-x_{0i})^2  \nonumber \\ 
&& +\frac{1}{3!} \sum_{i,j,k=1}^n a_{ijk} (\alpha)
(x_i-x_{0i})(x_j-x_{0j})(x_k-x_{0k})+\cdots \ , \label{expansion1}
\end{eqnarray}
where we assume that 
$\lambda_1(\alpha)$ becomes zero at $\alpha= \hat{\alpha}$ 
and takes negative value for $\alpha > \hat{\alpha}$, 
$\lambda_i (\alpha) < 0 \,(i=2 , \cdots , n)$ 
for $\alpha \geq \hat{\alpha}$ 
and $a_{111} (\hat{\alpha}) \neq 0$.
We would like to give an asymptotic formula 
of the integral~(\ref{Integral}) by extension of the method discussed in
Section 2.
The most direct method is to find a point $\tilde{\mbit{x}}$ 
so that $F(\mbit{x},\alpha)$ is expanded as 
\begin{eqnarray}
  F(\mbit{x},\alpha) &=& F(\tilde{\mbit{x}},\alpha) 
   + a_1 (\alpha) (x_1-\tilde{x}_1)
   + \frac{1}{2} \sum_{i=2}^n \tilde{\lambda}_i (\alpha) (x_i-\tilde{x}_i)^2
 \nonumber \\
&&+\frac{1}{3!} \sum_{i,j,k=1}^n \tilde{a}_{ijk}(\alpha)
(x_i-\tilde{x}_i)(x_j-\tilde{x}_j)(x_k-\tilde{x}_k)+\cdots \ .
\end{eqnarray}
If we can find out such a point $\tilde{\mbit{x}}$,
we obtain the asymptotic formula
\begin{eqnarray}
 I_n(\alpha) \sim (2 \pi i) e^{N F(\tilde{\mbit{x}},\alpha)}
\left( \frac{2 \pi}{N} \right)^{\frac{n-1}{2}} 
\left(\frac{2}{N \tilde{a}_{111}(\alpha)}  \right)^{\frac{1}{3}}
\left( \prod_{i=2}^n \left( -\tilde{\lambda_i} \right) 
\right)^{-\frac{1}{2}} \mbox{Ai}(\zeta') \ ,
\end{eqnarray}
where 
\begin{eqnarray} 
 \zeta (\alpha) \equiv
  - \left(\frac{2}{\tilde{a}_{111}(\alpha)} \right)^{\frac{1}{3}} 
  \tilde{\lambda}_1(\alpha) \ .
\end{eqnarray}
and we have only to rewrite this formula by $F(\mbit{x}_0,\alpha)$
and its derivatives at the saddle point $\mbit{x}_0$.

Although such a point $\tilde{\mbit{x}}$ always exacts,
it is difficult to exactly find out the point.
Fortunately, we do not need to derive explicitly the point.
As the same way in section 2, 
when $\alpha$ is in a neighborhood of $\hat{\alpha}$, 
a distance between $\tilde{\mbit{x}}$ and $\mbit{x}_0$, 
$|\tilde{\mbit{x}}-\mbit{x}_0|$, 
becomes very small.
Therefore we can expand $F(\tilde{\mbit{x}},\alpha)$ and its derivatives 
at the point $\tilde{\mbit{x}}$
around the saddle point $\mbit{x}_0$:
\begin{eqnarray}
 F(\tilde{\mbit{x}}, \alpha) &=& F(\mbit{x}_0,\alpha) 
+ \frac{1}{2} \lambda_1 (\tilde{x}_1-x_{01})^2 
+ \frac{1}{3!} a_{111} (\tilde{x}_1-x_{01})^3 + \cdots \ , \label{F0} \\
F_1^{(1)} (\tilde{\mbit{x}}, \alpha) &=&  \lambda_1 (\tilde{x}_1 - x_{01})
+\frac{1}{2} a_{111} (\tilde{x}_1 - x_{01})^2 + \cdots \ , \label{F11} \\
0=F_i^{(1)} (\tilde{\mbit{x}},\alpha) &=& \lambda_i (\tilde{x}_i-x_{0i}) 
+ \frac{1}{2} a_{i11} (\tilde{x}_1 - x_{01})^2 + \cdots \ ,  \,\,\, 
(i = 2, \cdots, n) \ , \label{F1i} \\
0=F_{11}^{(2)} (\tilde{\mbit{x}},\alpha) &=&  \lambda_1 
+ a_{111} (\tilde{x}_1 - x_{01}) + \cdots \ , \label{F211} \\
F_{1i}^{(2)} (\tilde{\mbit{x}},\alpha) &=& a_{i11}(\tilde{x}_1 - x_{01}) 
+ \cdots = F_{i1}^{(2)} (\tilde{\mbit{x}},\alpha) \ , \,\,\, 
(i = 2, \cdots, n) \ , \\
F_{ij}^{(2)} (\tilde{\mbit{x}},\alpha) &=& \lambda_i \delta_{ij} 
+ a_{1ij} (\tilde{x}_1 - x_{01}) + \cdots \ , \,\,\, 
(i, j = 2, \cdots, n) \ , \label{F2ij} \\
F_{111}^{(3)} (\tilde{\mbit{x}},\alpha) &=& a_{111} + \cdots \ , \label{F3111}
\end{eqnarray}
where we have used the abbreviations
\begin{eqnarray}
 F_{i_1 i_2 \cdots i_k}^{(k)} (\tilde{\mbit{x}},\alpha) 
= \left. \frac{\partial^k F(\mbit{x},\alpha)}
{\partial x_{i_1} \partial x_{i_2} \cdots \partial x_{i_k}} 
\right|_{\mbit{x}=\tilde{\mbit{x}}} \ .  
\end{eqnarray}
We assume that $\lambda_1(\alpha) \sim O (1/N^{\frac{1}{3}})$
and that $N$ is sufficiently large.
As in an integral with one variable, Eqs.~(\ref{F1i}), (\ref{F211}) give
\begin{eqnarray}
 \tilde{x}_1 - x_{01} \approx -\frac{\lambda_1}{a_{111}} \ , \,\,\, 
\tilde{x}_i -x_{0i} \approx -\frac{a_{i11}}{2 \lambda_i a_{111}^2} \lambda_1^2
\,\,\, (i=2, \cdots, n) \ , 
\end{eqnarray}
and Eqs.~(\ref{F0}), (\ref{F11}), (\ref{F2ij}) and (\ref{F3111}) 
are approximately given by 
\begin{eqnarray}
 F(\tilde{\mbit{x}},\alpha) &\approx& F (\mbit{x}_0,\alpha) 
- \frac{1}{3} \frac{(-\lambda_1)^3}{a_{111}^2} \ , \label{F0a}\\
F_1^{(1)} (\tilde{\mbit{x}},\alpha) &\approx&  
- \frac{1}{2} \frac{(-\lambda_1)^2}{a_{111}} \ , \\ 
F_{ij}^{(2)} (\tilde{\mbit{x}},\alpha) &\approx& 
 \lambda_i(\alpha) \delta_{ij} \ , \,\,\, 
(i, j = 2, \cdots, n) \ , \\
F_{111} (\tilde{\mbit{x}},\alpha) &\approx& a_{111}(\alpha) \label{F3111a} \ ,
\end{eqnarray}
respectively.
Although the terms 
$F_{1i}^{(2)}(\tilde{\mbit{x}},\alpha) \,\,\, (i = 2, \cdots, n)$ 
remain finite, 
these turns out to be suppressed as higher order terms.
The use of Eqs.~(\ref{F0a})-(\ref{F3111a}) gives  
\begin{eqnarray}
 I_n(\alpha , N) &\sim& \int d^n x \exp \Bigg[N 
\Bigg\{ F(\mbit{x}_0, \alpha) - \frac{1}{3} \frac{(-\lambda_1)^3}{a_{111}^2} 
\nonumber \\
&&- \frac{1}{2} \frac{(-\lambda_1)^2}{a_{111}} (x_1-\tilde{x}_1) 
+ \frac{a_{111}}{3!} (x_1-\tilde{x}_1)^3 
+ \frac{1}{2} \lambda_i (x_i-\tilde{x}_i)^2 
+ \cdots \Bigg\} \Bigg] \ .
\end{eqnarray}
Making changes of variables 
\begin{eqnarray}
&& x_1-\tilde{x}_1 = \frac{1}{N^{\frac{1}{3}}} 
\left( \frac{2}{a_{111}} \right)^{\frac{1}{3}} t_1 \ , \ 
x_i-\tilde{x}_i = \frac{1}{N^{\frac{1}{2}}} t_i \ , \,\,\,
(i = 2, \cdots, n) \ , \\ 
&& \zeta (\alpha) =\left( \frac{(-\lambda_1(\alpha))^3}{2 a_{111}^2(\alpha)}  
\right)^{\frac{2}{3}} \ ,
\end{eqnarray}
we obtain  
an asymptotic formula of the integral~(\ref{Integral}),
\begin{eqnarray}
 I_n (\alpha,N) &\sim& e^{N F(\mbit{x}_0 , \alpha) }
\left( \frac{2 \pi}{N} \right)^{\frac{n}{2}}
\frac{1}{\left[ \det \left\{-F^{(2)}(\mbit{x}_0, \alpha) \right\} 
\right]^{\frac{1}{2}}} (2 \sqrt{\pi} i) \zeta^{'\frac{1}{4}}
e^{-\frac{2}{3} \zeta^{' \frac{3}{2}} } \mathrm{Ai}(\zeta') \ ,
\label{asymptotic2}
\end{eqnarray}
where we have used $\zeta' = N^{\frac{2}{3}} \zeta$.
The divergence of the term 
$\det \left\{-F^{(2)}(\mbit{x}_0, \alpha) \right\}$ 
at $\alpha=\hat{\alpha}$ is suppressed by the term
$\zeta^{'\frac{1}{4}}$.
Thus we can derive the asymptotic formula without the another saddle
point which coincides with the saddle point $\mbit{x}_0$ 
at $\alpha=\hat{\alpha}$.
Also, in the same way as section 2, 
it is not difficult to show that the asymptotic formula~(\ref{asymptotic2}) 
coincides with the uniform asymptotic expansion derived 
by N.~Bleistein \cite{rf:B} in the neighborhood of $\hat{\alpha}$.

Finally, we can simply comment about the mean field approximation.
As same way as the asymptotic formula~(\ref{asymptotic}),
the form of the asymptotic formula~(\ref{asymptotic2}) 
is the form attached the extra term to the leading order of the 
asymptotic expansion based on the WKB approximation. 
Therefore the mean field approximation based on 
the asymptotic formula~(\ref{asymptotic2}) coincides with the one based
on the WKB approximation.
This fact gives the important result that  
the mean field approximation of the model with caustics 
is justified in the sense of the formula~(\ref{asymptotic2}).

\section{Conclusion and Discussion}

In this paper, we derive the asymptotic formula for the integral with
caustics, which we can utilize when the integral has 
only one saddle point $\mbit{x}_0$ or at least 
when we do not know whether the another saddle point, 
which coincides with $z_0$ at $\alpha=\hat{\alpha}$, exists.
This asymptotic formula 
coincides with the uniform asymptotic expansion 
in the neighborhood of $\hat{\alpha}$.  
Of course, if we can find two saddle points, we should utilize the
method presented by C.~Chester $\textit{et al.}$ \cite{rf:CFU, rf:F, rf:U} 
and N.~Bleistein \cite{rf:B} so that we can obtain the uniform
asymptotic expansion.
Also, we show the important result that 
the mean field approximation of the model with caustics 
is justified in the sense of the asymptotic formula whose 
result converges.
This result guarantees the validity for the mean field approximation 
of the NJL model. 
      
Our next aim is to apply this result to quantum field theoretical
models with caustics.
We think that carrying out this is very important: 
in reference~\cite{rf:KS}, 
we calculated the effective potential and Gap equation of the NJL model.
Then in order to suppress the infrared divergence which 
occurs from the loop diagrams of the massless Nambu-Goldstone boson, 
we introduced a fermion mass in advance.
According to reference~\cite{rf:KS2}, 
the result of the WKB approximation becomes bad not only 
at $\alpha = \hat{\alpha}$ but also in the neighborhood of
$\hat{\alpha}$, 
whereas the result of the approach mentioned in section 2
gives very accurate approximation in this region.  
Therefore if the introduced mass is sufficiently small 
in comparison with the typical energy scale, 
there is possibility that 
the asymptotic formula derived in this paper
gives better result than the WKB approximation.  
In this sense, 
re-estimate for quantum (= loop) effects of 
auxiliary fields in the effective potential and gap equation 
of the NJL model is very interesting.
Also, in reference~\cite{rf:OH},
A.~Osipov $\textit{et al.}$ first calculated the quantum
correction of the gap equation and effective potential 
of the auxiliary fields for the NJL model with the 't Hooft
interactions and dealt with the change of the vacuum state 
due to it.
Then they suggested that including the quantum correction causes
caustics at some values of the coupling constants.
The use of the asymptotic formula to these caustics 
is interesting issue.

\vspace{10mm}
\noindent{\Large \bf Acknowledgements} \\
He would like to thank Koji Harada 
for reading this manuscript and
making useful suggestions.

\end{document}